\documentclass{llncs}
\usepackage{url}
\usepackage{graphicx}
\begin{document}

\title{Thirteen Years of Mining Software Repositories (MSR) Conference - What is the Bibliography Data Telling Us?}
\author{Lov Kumar\inst{1} \and Ashish Sureka\inst{2}}
\institute{
NIT Rourkela, India,\\
\email{lovkumar505@gmail.com}\\
\and
ABB Corporate Research, India,\\
\email{ashish.sureka@in.abb.com}\\
}
\maketitle

\begin{abstract}
The Mining Software Repositories (MSR) conference is a reputed, long-running and  flagship conference in the area of Software Analytics which has successfully completed more than one decade as of year $2016$. We conduct a bibliometric and scientific publication mining based study to study how the conference has evolved over the recent past $13$ years (from $2004$ to $2007$ as a workshop and then from $2008$ to $2016$ as a conference). Our objective is to perform an examination of the state of MSR so that the MSR community can identify strengths, areas of improvements and future directions for the conference. 
\end{abstract}

\section{Introduction}
\noindent\textbf{Research Motivation and Aim:} The Mining Software Repositories (MSR) conference is an annual conference started as a one-day workshop in the year $2004$ (co-located with International Conference on Software Engineering ICSE 2004 at Edinburgh, UK) with the aim of brining software analytics researchers and practitioners from both university and industry around the world to exchange research results and ideas. MSR $2004$ (Edinburgh, UK) was the first edition of the event and the annual event completes $13$ editions in $2016$ (Texas, USA). We believe that a reflection of the $13$ years of MSR is important for the MSR community to learn from its history and gain insights for further improving the visibility and quality of the conference.  Our research motivation is to investigate answers to questions such as: how the conference has evolved over the past $13$ years and what is the current status, what is the quality of the conference based on several key performance indicators, what improvements can be made and to what extent MSR is meeting its desired objectives. Our research aim is to systematically and scientifically explore and examine the state of MSR across various aspects of the conference. To the best of our knowledge, the study presented in this paper is the first in-depth examination of the state of MSR which we believe is important for the MSR community to understand its development, evolution and identify future directions.\\
\begin{table}[htbp]
	\caption{Number of Papers Submitted (SUB), Types of Paper accepted (2004-2007)}
	\label{msrwork}
	\centering
		\renewcommand{\arraystretch}{1.1}
		\resizebox{8.0cm}{!}{
	\begin{tabular}{|c|c|p{6.5cm}|} 
		\hline
		\textbf{Year} & \textbf{SUM} & \textbf{Accepted} \\ \hline
	2004 & 38 & 	26 (All 4 Page)	 \\ \hline						
	2005  &	38 &	11 (5 Page Regular), 11(5 Page Light Talk) \\ \hline					
	2006 &	45 & 16 (7 Page Full),	12 (3 Page Short),  11 (2 Page Challenge Report) \\ \hline		
	2007 &	52 & 16	(8 Page Full),	12(4 pages Short), 3 (2 Page challenge Report), 3 (Prediction) \\ \hline
\end{tabular}}
\end{table}	
\begin{table*}[htbp]
	\caption{Number of Full Papers Submitted (SFP), Number of Full Papers Accepted (AFP), Number of Short Papers Submitted (SSH), Number of Short Papers Accepted (ASH),  Number of Data Showcase Submitted (SDS), Number of Data Showcase Accepted (ADS),   Number of MSR Challenge Papers Submitted (SCH), Number of MSR Challenge Papers Accepted (ACH), and Acceptance Rate (AR) }
	\label{msrconf}
	\centering

		\begin{tabular}{|c|c|c|c|c|c|c|c|c|c|c|c|c|} 
			\hline
		
		\textbf{Year} & \textbf{SFP} & \textbf{AFP} & \textbf{AR} & 	\textbf{SSH}	& \textbf{ASP} & \textbf{AR} &	\textbf{SDS} &	\textbf{ADS} & \textbf{AR} &	\textbf{SCH}	& \textbf{ACP} &\textbf{AR}  \\ \hline
		\textbf{2016} & 103 & 36 & 34.95\% & 30 & 6 & 20.00\% & 13 & 7 & 53.85\% & 24 & 10 & 41.67\%  \\ \hline
	   	\textbf{2015} & 106 & 32 & 30.19\% & 20 & 10 & 50.00\% & 25 & 16 & 64.00\% & 21 & 14 & 66.66\%  \\ \hline
	   	\textbf{2014} & 85 & 29 & 34.12\% & 27 & 10 & 37.04\% & 22 & 15 & 68.18\% & 19 & 9 & 47.37\%  \\ \hline
	   	\textbf{2013} & 86 & 29 & 33.72\% & 22 & 5 & 22.73\% & 27 & 15 & 55.55\% & 29 & 12 & 41.38\%  \\ \hline
	   	\textbf{2012} & 64 & 18 & 28.13\% & 22 & 12 & 54.54\% & NA & NA & NA & 17 & 6 & 35.3\%  \\ \hline
	   	\textbf{2011} & 61 & 20 & 32.79\% & 17 & 6 & 35.29\% & NA & NA & NA & 6 & 5 & 83.33\%  \\ \hline
	   	\textbf{2010} & 51 & 15 & 29.41\% & 16 & 5 & 31.25\% & NA & NA & NA & 9 & 6 & 66.66\%  \\ \hline
	   	\textbf{2009} & 47 & 12 & 25.53\% & 18 & 10 & 55.56\% & NA & NA & NA & 9 & 5 & 55.55\%  \\ \hline
	   	\textbf{2008} & 21 & 8 & 38.10\% & 21 & 14 & 66.67\% & NA & NA & NA & - & 5 & -  \\ \hline
		
		\end{tabular}
	\end{table*}
	
\noindent\textbf{Related Work:} Robles et al. review all papers published in the proceedings of MSR from $2004$ to $2009$. They analyze the papers that contained any experimental analysis of software projects for their potentiality of being replicated \cite{robles2010}. Hemmati et al. review $117$ full papers published in the MSR proceedings between $2004$ and 2012 \cite{hemmati2013}. They extract $268$ comments from $117$ papers, categorize them using a grounded theory methodology and create high-level themes \cite{hemmati2013}. Tripathi et al. study $5$ years of research papers published in MSR series of conferences ($2010$-$2014$) and present insights on the number of studies using solely Open Source Software (OSS) data or solely (Closed Source Software) CSS data or both OSS and CSS data \cite{tripathi2015}. They also count the number of papers published by authors solely from Universities, solely from Industry and from both University and Industry \cite{tripathi2015}.
\section{Bibliometric Analysis \& Results}
\noindent\textbf{Paper Acceptance Rate:} We download all the papers published in $13$ years of MSR. We also download the message from the General and Program Chairs which are published as part of the conference proceedings. Table \ref{msrwork} and \ref{msrconf} displays the number and types of papers submitted across various tracks from the year $2004$ to $2016$. We extract the information about submitted and accepted papers from the message from the conference chairs and the PDF files of the papers which we were able for download. Table \ref{msrconf} reveals that the number of full or regular papers submitted increased from $21$ in $2008$ to $103$ in $2016$. The acceptance rate for full papers varies from a minimum of $25.53\%$ to a maximum of $38.10\%$ during a period of $2008$ to $2016$. Table \ref{msrconf} shows that MSR invites a variety of submissions in addition to regular papers such as short papers, data showcase and data challenge. \\ 								
\begin{table}[htbp]
	\caption{Descriptive Statistics for MSR $2004$ to $2016$ Google Scholar Citations}
	\label{citetable}
	\centering
	\begin{tabular}{|c|c|c|c|c|c|c|c|} 
		\hline
		\textbf{Year} & \textbf{Min.} & \textbf{Max.} & \textbf{Mean} & \textbf{Median} & \textbf{Sum} \\ \hline
2004 & 0 & 253 & 47.38 & 27.5 & 1232 \\ \hline
2005 & 17 & 487 & 69.91 & 46 & 1538 \\ \hline
2006 & 1 & 440 & 46.15 & 29 & 1800 \\ \hline
2007 & 3 & 242 & 53.76 & 39 & 1774 \\ \hline
2008 & 4 & 120 & 37.22 & 24 & 1005 \\ \hline
2009 & 0 & 160 & 38.85 & 26 & 1049 \\ \hline
2010 & 5 & 197 & 36.38 & 25 & 946 \\ \hline
2011 & 1 & 111 & 28.45 & 27 & 882 \\ \hline
2012 & 2 & 93 & 22.36 & 18 & 805 \\ \hline
2013 & 0 & 109 & 20.20 & 14 & 1232 \\ \hline
2014 & 0 & 81 & 12.90 & 7 & 813 \\ \hline
2015 & 0 & 15 & 2.46 & 1 & 177 \\ \hline
2016 & 0 & 1 & 0.05 & 0 & 3 \\ \hline
ALL & 0 & 487 & 25.39 & 12 & 13256 \\ \hline
\end{tabular}
\end{table}	
\begin{table*}[htbp]
	\caption{Top $10$ Most Cited MSR $2004$ to $2016$ Papers Based on Google Scholar Metrics (Citations Metrics Collected on 15 June 2016)}
	\label{top10citedpaper}
	\centering
	\renewcommand{\arraystretch}{1.1}
		\begin{tabular}{|c|c|p{5.5cm}|p{1.5cm}|l|c|} 
			\hline
			\textbf{Rank} & \textbf{Year} & \textbf{Paper Title} & \textbf{First Author} & \textbf{Country} & \textbf{Citations} \\ \hline
		1 & $ 2005 $ & When do changes induce fixes? & Jacek Sliwerski & Germany & 487 \\ \hline
		2 & $ 2006 $ & Mining email social networks. & Christian Bird & USA & 440 \\ \hline
		
		3 & $ 2004 $ & Preprocessing CVS Data for Fine-Grained Analysis  & Thomas Zimmermann & USA & 253 \\ \hline
		
		4 & $ 2007 $ & How Long Will It Take to Fix This Bug? & Cathrin WeiB & Germany & 242 \\ \hline
		
		5 & $ 2010 $ & An extensive comparison of bug prediction approaches. & Marco D'Ambros & Switzerland & 197 \\ \hline
		
		6 & $ 2004 $ & The Perils and Pitfalls of Mining SourceForge  & James Howison & USA & 196 \\ \hline
		
		7 & $ 2006 $ & MAPO: mining API usages from open source repositories. & Tao Xie & USA & 173 \\ \hline
		
		8 & $ 2009 $ & The promises and perils of mining git. & Christian Bird & USA & 160 \\ \hline
		
		9 & $ 2005 $ & Understanding source code evolution using abstract syntax tree matching. & Iulian Neamtiu & USA & 155 \\ \hline
		
		10 & $ 2004 $ & Applying Social Network Analysis to the Information in CVS Repositories  & Luis Lopez-Fernandez & Spain & 155 \\ \hline
		
		\end{tabular}
	\end{table*}
\begin{figure}[htbp]
		\centering
		\includegraphics[width=12.0cm, height=4cm]{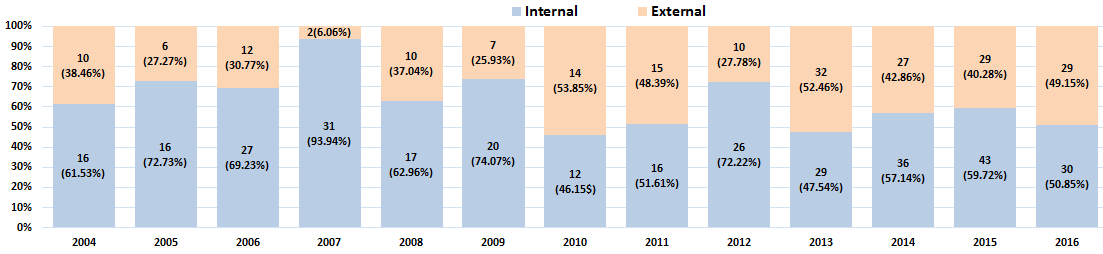}
		\caption{Stacked Bar Chart Indicating the Percentage Distribution of Internal and External Collaboration}
		\label{internalexternal}
	\end{figure}
\begin{figure*}[th]
	\centering
	\includegraphics[width=12.0cm, height=4cm]{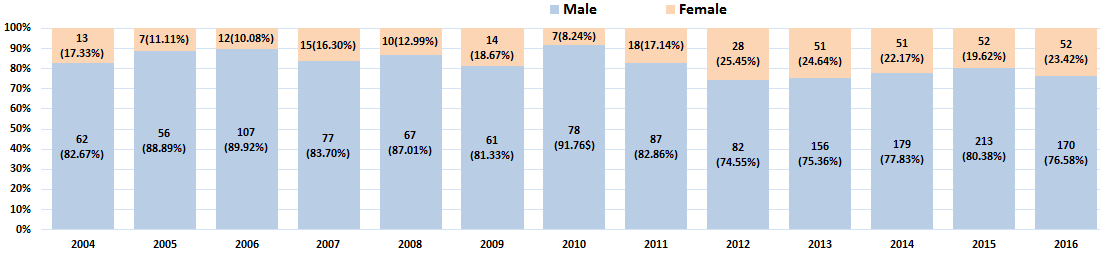}
	\caption{Percentage of Male and Female Authors (Gender Imbalance)}
	\label{gender}
\end{figure*}
\begin{figure*}[h]
	\centering
	\includegraphics[width=12.0cm, height=4cm]{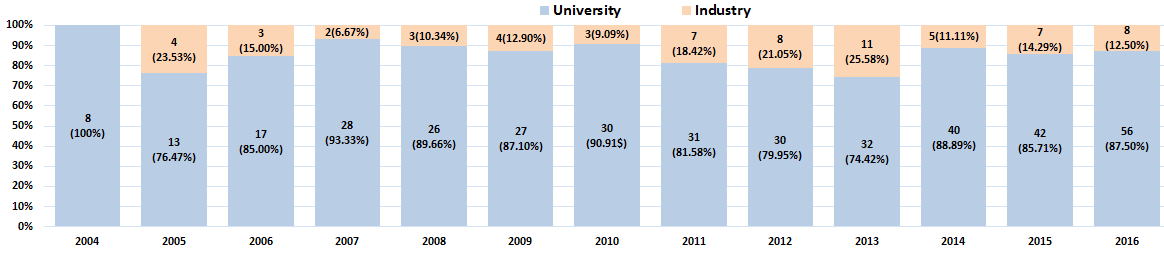}
	\caption{Percentage Distribution of Program Committee Members across Industry and University}
	\label{pcunivind}
\end{figure*}
\begin{table}[h]
	\centering
	\caption{Number and Percentage of Research Track Papers having All Authors from University (AU), All Authors from Industry (AI) and Authors from both University and Industry (UI)}
	\label{uicollab}
	\begin{tabular}{|l|*{5}{c|}r}
		\hline
		\textbf{Year} & \textbf{NUM} & \textbf{AU} & \textbf{AI} & \textbf{UI} \\ \hline
		2004 & 26 & 20 (76.92\%) & 1 (3.85\%) & 5 (19.23\%)  \\ \hline
		2005 & 22 & 20 (90.91\%) & 1 (4.55\%) & 1 (4.55\%)  \\ \hline
		2006 & 39 & 32 (82.05\%) & 2 (5.13\%) & 5 (12.82\%)  \\ \hline
		2007 & 33 & 28 (84.85\%) & 5 (15.15\%) & 0 (0\%)  \\ \hline
		2008 & 27 & 22 (81.48\%) & 1 (3.7\%) & 4 (14.81\%)  \\ \hline
		2009 & 27 & 21 (77.78\%) & 1 (3.7\%) & 5 (18.52\%)  \\ \hline
		2010 & 26 & 21 (80.77\%) & 0 (0\%) & 5 (19.23\%)  \\ \hline
		2011 & 31 & 26 (83.87\%) & 2 (6.45\%) & 3 (9.68\%)  \\ \hline
		2012 & 36 & 31 (86.11\%) & 3 (8.33\%) & 2 (5.56\%)  \\ \hline
		2013 & 61 & 48 (78.69\%) & 2 (3.28\%) & 11 (18.03\%)  \\ \hline
		2014 & 63 & 51 (80.95\%) & 2 (3.17\%) & 10 (15.87\%)  \\ \hline
		2015 & 72 & 57 (79.17\%) & 5 (6.94\%) & 10 (13.89\%)  \\ \hline
		2016 & 59 & 51 (86.44\%) & 2 (3.39\%) & 6 (10.17\%)  \\ \hline
		ALL & 522 & 428 (81.99\%) & 27 (5.17\%) & 67 (12.84\%)  \\ \hline
	\end{tabular}
\end{table}

\noindent\textbf{Citation Based Impact:} The h5-index for MSR on $16$ July $2016$ is $34$. Google Scholar defines h5-index as "h5-index is the h-index for articles published in the last 5 complete years. It is the largest number h such that h articles published in 2011-2015 have at least h citations each". The h5-median for MSR on $16$ July $2016$ is $46$. Google Scholar defines h5-median as "h5-median for a publication is the median number of citations for the articles that make up its h5-index". 

Table \ref{citetable} shows the descriptive statistics for MSR $2004$ to $2016$ Google Scholar Citations as on $15$ June $2016$. Table \ref{citetable} reveals that in $13$ years MSR papers have received a total of $13256$ citations. It is interesting to note that even when MSR was a workshop from $2004$ to $2007$ and small in scale, still the total number of citations of all the published papers are more than $1500$ for every year. Table \ref{top10citedpaper} shows the top $10$ most cited MSR $2004$ to $2016$ papers Based on Google Scholar Metrics (Citations Metrics Collected on 15 June 2016).  The h5-index of the top tier conference in SE (ICSE) is $63$. The h5-index of MSR with respect to ICSE and the data in Table \ref{top10citedpaper} and \ref{citetable} shows that MSR papers have high citation impact. Table \ref{top10citedpaper} shows that the most cited paper in MSR has $487$ citations and the Top $10$ most cited papers have more than $150$ citations. \\

\noindent\textbf{University-Industry Collaboration:} Joint authorship in scientific papers is an evidence of collaboration and interaction between researchers as well as institutions. Our objective is to study university-industry collaboration and knowledge flow between the two types of institutions. Table \ref{uicollab} displays the data on university-industry collaboration. Table \ref{uicollab} reveals that the percentage of joint university-industry papers (research track) varies from a minimum of $0\%$ (in the year $2007$) to a maximum of $19.23\%$ (in the year $2004$ and $2010$). Table \ref{uicollab} reveals that out of the $522$ research track papers published in MSR in $13$ years, the number and percentage of papers involving a university-industry collaboration is $67$ and $12.84$ respectively. \\
	
\noindent\textbf{Internal-External Collaboration :} We investigate the nature and scale of collaboration in MSR papers from the perspective of internal or external collaboration. Internal collaboration is a form of collaboration in which all the co-authors in a paper (single or multiple-authors) are affiliated to one Institution only. External collaboration is defined as a form of collaboration which involves participation of two or more institutions (irrespective of whether the organizations involved are industry or university) in the production of the scientific output and the paper. Figure \ref{internalexternal} displays a stacked bar chart indicating the percentage distribution of internal and external collaboration. Figure \ref{internalexternal} reveals a good percentage of external collaboration. The percentage of papers having external collaboration varies from a minimum of $6.06\%$ to a maximum of $53.85\%$.	\\

\noindent\textbf{Gender Imbalance in Authorship:} Agarwal et al. conduct an analysis of women in computer science research by analyzing author data from $81$ conferences including $11$ conferences in software engineering \cite{swati2016}. Their experimental dataset consists of DBLP bibliography entries from the year $2000$ to $2015$. Their results reveal that $79\%$ of the authors in the bibliography dataset consisting of $11$ conferences and $16$ years are male whereas $21\%$ authors are women authors \cite{swati2016}. We use the Genderize.io\footnote{\url{https://genderize.io/}} API to determine the gender of all the authors in our dataset. Figure \ref{gender} displays a stacked-bar chart showing the percentage of male and female authors every year from $2004$ to $2016$. Figure \ref{gender} reveals a gender imbalance in authorship. The percentage of female authors varies from a minimum of $8.24\%$ in the year $2010$ to a maximum of $25.45\%$ in the year $2012$. We observe that the percentage of female authors is less than $20\%$ for $9$ out of $13$ years. \\

\noindent\textbf{Program Committee Characteristics:} The size of the program committee should be according to the number of papers normally received by the conference so that the workload of the program committee members is reasonable or moderate. We extract the size of the program committee (for both the research and industry track) from the MSR conference proceedings. Table \ref{programcountry} shows that the number of program committee members varies from a minimum of $8$ in year $2004$ (the first edition when MSR was a workshop) to a maximum of $64$ in $2016$. The number of papers submitted at MSR in the past two years ($2015$ and $2016$) is in the range of $10$ to $110$ and hence the distribution of workload to the committee members is moderate. 

Diversity of institution, technical area of expertise and country is an important selection criteria for selecting program committee member and is an indicator of the quality of a conference. We extract the country of every program committee member and compute the number of different countries. Table \ref{programcountry} shows that MSR program committee is diverse and inclusive in-terms of the number of countries. For example, in the year $2016$, there were $64$ program committee members from $33$ different countries (a diversity score of $51.56\%$). In year $2015$, there were $49$ members from $27$ countries. 

Annual churn and rotation of program committee members is essential for making sure that there is diversity, inclusiveness and cross-section of topic expertise, institution and geographical area. Inviting new program committee members and making space for them by rotating-off program committee members who have served for $2-3$ years are normal guidelines for conferences. We compute the yearly churn in the program committee for MSR $2004$ to MSR $2016$. Table \ref{programcountry} shows that in the year $2015$ there were a total of $49$ program committee members out of which $29$ ($59.18\%$) were new and $20$ were repeated from the previous years We observe that the highest churn was in the year $2016$ ($70.31\%$) and the lowest was in the year $2008$ ($34.48\%$). It is an important guideline for program committee chairs who lead the program committee member selection and invitation process to have a balanced representation from both industry and academia. We extract the affiliation of each program committee member and determine whether the member belongs to an industry or university.  

Figure\ref{pcunivind} reveals an imbalance between industry and academia and is skewed towards university. The percentage of program committee members from university varies from a minimum of $74.42\%$ to a maximum of $100\%$. We observe that for $10$ out of $13$ years, the percentage of program committee members from industry is less than $20\%$.
\begin{table}[h]
	\centering
	\caption{Program Committee Characteristics}
	\label{programcountry}
	\begin{tabular}{|l|*{3}{c|}r}
		\hline
		\textbf{Year} & \textbf{NUM} & \textbf{ Country} & \textbf{NEW} \\ \hline
	2004 & 8 & 5 (62.5\%) &   \\ \hline
	2005 & 17 & 5 (29.41\%) & 11 (64.71\%)  \\ \hline
	2006 & 20 & 8 (40\%) & 11 (55\%)  \\ \hline
	2007 & 30 & 15 (50\%) & 18 (60\%)  \\ \hline
	2008 & 29 & 10 (34.48\%) & 10 (34.48\%)  \\ \hline
	2009 & 31 & 10 (32.26\%) & 17 (54.84\%)  \\ \hline
	2010 & 33 & 11 (33.33\%) & 16 (48.48\%)  \\ \hline
	2011 & 38 & 12 (31.58\%) & 23 (60.53\%)  \\ \hline
	2012 & 38 & 13 (34.21\%) & 17 (44.74\%)  \\ \hline
	2013 & 43 & 15 (34.88\%) & 17 (39.53\%)  \\ \hline
	2014 & 45 & 15 (33.33\%) & 28 (62.22\%)  \\ \hline
	2015 & 49 & 27 (55.1\%) & 29 (59.18\%)  \\ \hline
	2016 & 64 & 33 (51.56\%) & 45 (70.31\%)  \\ \hline
	\end{tabular}
\end{table}	
\section{Conclusion}
We conclude that MSR is successfully meeting its desired objective as it is able to attract a good number of papers from different parts of the world both from industry and academia. The acceptance rate demonstrates that MSR is a moderately selective conference. The citation impact of the conference is high indicating that MSR is maintaining its status as a Tier $2$ conference. The papers published in MSR demonstrates both university-industry collaboration as well as external collaboration. The program committee of MSR is diverse both from the perspective of representations from industry and academia and from different countries. There is a healthy program committee and author churn which indicates that the conference is broad and open. MSR authorship indicates a gender imbalance and low percentage of women authors.
\bibliographystyle{acm}
\bibliography{sigproc}
\end{document}